\documentclass[aps,prb,twocolumn,superscriptaddress,showpacs]{revtex4}
\usepackage{amsmath}

\begin{document}

\title{Polarizability calculation of vibrating
nanoparticles for intensity of low frequency Raman scattering}

\author{Daniel B. Murray}
\email{daniel.murray@ubc.ca}
\author{Caleb H. Netting}
\email{calebnetting@gmail.com}
\author{Robin D. Mercer}
\affiliation{Mathematics, Statistics and Physics Unit,
University of British Columbia Okanagan \\
3333 University Way, Kelowna, British Columbia, Canada V1V 1V7}

\author{Lucien Saviot}
\email{lucien.saviot@u-bourgogne.fr}

\affiliation{Laboratoire de Recherche sur la R\'eactivit\'e des Solides,
UMR 5613 CNRS - Universit\'e de Bourgogne\\
9 avenue A. Savary, BP 47870 - 21078 Dijon - France}

\date{\today}

\begin{abstract}
A new numerical method is introduced for calculating the
polarizability of an arbitrary dielectric object with position
dependent complex permittivity.  Three separate numerical
approaches are provided to calculate the dipole moment of a
nanoparticle embedded in a dielectric matrix in the presence
of an applied electric field.  Numerical tests confirm the
accuracy of this method when applied to several cases for
which an exact solution is available.  This method is especially
well suited for the calculation of absolute Raman scattering
intensities due to acoustic phonons in metallic and dielectric
nanoparticles embedded in transparent matrices.

\end{abstract}

\maketitle

Key words: Nanoparticle, Raman intensity, polarizability,
dipole moment, permittivity

\section{Introduction}
\label{secI}

Spectral features of low frequency Raman scattering from
nanoparticles (NP) can be explained in terms of the
vibrational frequencies of acoustic phonons which
are confined in those NP.\cite{duval86}
Calculation of the displacement fields of the modes
permits prediction of the Raman selection rules.
Variation of Raman intensity when parallel or
perpendicular polarization is selected can also be
understood.

Past theoretical work on the intensity of Raman
scattering from NP\cite{gersten80,montagna95,bachelier04}
has not produced quantitative estimates of scattering
intensities.  However, key qualitative features, such
as the distinction between NP volume and surface
mechanisms,\cite{bachelier04} have been pointed out.
In the NP volume mechanism, deformation potential coupling
modulates the bulk dielectric response.\cite{delfatti00}
In the surface coupling mechanism, changes of the
NP's size or shape modulate the NP's polarizability.\cite{gersten80}

Calculation of the absolute intensity of low
frequency Raman scattering from an NP requires the
values of the polarizability matrix, $\alpha$, of the NP,
in particular the modulation of the polarizability with
time due to the acoustic phonon degrees of freedom.  Our
approach incorporates both the NP volume and surface
coupling mechanisms.  In addition, our general approach
includes an additional qualitative mechanism which has
not been previously pointed out:  Variations of the
density of the matrix material surrounding the NP will
lead, through deformation potential coupling, to a new
matrix volume mechanism.  

It is a reasonable approximation in many cases to hope to
determine the polarizability of an object in terms of a local
permittivity function, $\epsilon({\bf r})$ such that
${\bf D}({\bf r})$=$\epsilon({\bf r}){\bf E}({\bf r})$.
Our discussion only applies to this case.

In some kinds of NPs this approach will not work.  It is not
always the case that the dielectric response of an object can be
described in terms of a local permittivity function.  An example
of this is when the Raman scattering of a NP is dominated by acoustic
phonon modulation of electron hole excitons, as in semiconductor
NPs when the Raman laser is close to the exciton energies.  In
addition, local dielectric response cannot be expected for
distance scales comparable to the Thomas-Fermi length.  Such
situations will not be considered further in what follows.

What is needed is a method which can handle:
(1) small variations in shape of an object;
(2) complex permittivity;
(3) permittivity which is an arbitrary function of position;
(4) non-spherical shape;
(5) permittivity with a large magnitude.

Quite a number of methods are available to calculate
the polarizability of an object.  However, none of
these methods are suitable for the requirements of
low frequency Raman scattering from NPs.

The exact solution due to Mie\cite{mie1908} is for homogeneous
spherical objects only.  The most general method is Finite
Difference Time Domain (FDTD).\cite{taflove95}  However, FDTD
cannot handle very small changes in the object since it must
use a coarse grid of spatial points.  The discrete dipole
approximation (DDA)\cite{purcellAPHJ73,drainAPHJ88,kellyJPCB03}
requires a mesh to approximate the object, and is not suitable
to reflect small changes in shape due to vibration.  The
dipole-induced-dipole (DID) method\cite{montagna95} is
applicable only when the permittivity is small.

No presently available numerical method meets all of these
criteria.  This paper introduces a new method which satisfies
all of these requirements.  The application of this method will
permit quantitative estimates of the intensity of low frequency
Raman scattering due to acoustic phonons in NPs.

This paper is restricted only to the problem of
calculating the polarizability tensor for a given
static configuration of a NP.  Repetition
of this method for a sequence of configurations
associated with the motion of an acoustic phonon
will allow the modulation of the polarizability
tensor to be determined.  This leads directly to
the scattered Raman intensity, which is the ultimate
justification for this work.

NPs are very small compared to the wavelength of the laser light
used to excite them in Raman and Brillouin light scattering
experiments.  Thus, at an instant of time the electric field in
the region enclosing a NP may be regarded as a uniform static
electric field.  The NP has a dielectric constant which differs
from that of the surrounding glass matrix.  For this reason
there will be a spatial variation of the electric field in the
interior and vicinity of the NP.  This electric field induces
polarization and consequently bound charge on the NP which leads
to a dipole moment.  It is this electric dipole that oscillates
so as to radiate, emitting Rayleigh scattered light (with the
same wavelength as the incident light).  Acoustic vibrations of
the NP lead to variations in the dielectric response of the NP.
Their frequency is very low compared to the laser.
As the NP slowly vibrates, its dielectric response leads to
Raman or Brillouin scattered light whose frequency is slightly
above or below that of the incident laser.

An electrostatic (also called ``quasistatic'') description
of a system is justifiable if the characteristic frequencies
applied are much less than the speed of light divided by the
diameter of the system.  For a NP, this intrinsic frequency
would be roughly $10^{17}$~Hz, whereas the frequency of light is
below $10^{15}$~Hz.  Consideration of retardation effects is not
important if we are considering small NPs.  In addition, note
that the NP oscillates very slowly compared to the incident
electric field.

To help understand the physical situation in question, consider
a single instant in time. Consider also a region surrounding
the NP over which the incident electric field is
approximately constant. The incident electric field ${\bf E}_{inc}$
has components $E_{incx}$, $E_{incy}$, and $E_{incz}$.
If the region under consideration contained only glass without a NP,
then the electric field would be constant within the region.
The effect of the variation of the dielectric response of the NP
material relative to that of the glass is that the electric
field varies in and close to the NP.

In the following sections, a numerical method is introduced for
determining the resulting electric field associated with the NP
as well as the dipole moment of the NP.  This can be used to
find the polarizability of the NP.

\section{Dipole Radiation}
\label{secIb}

Before continuing on to the main point of this paper, which is
the calculation of the polarizability of a NP, we will here
present the key relationships that will permit these results
to be applied to the problem of the calculation of the intensity
of Raman scattering.

First, note that, for a small object, the radiation emitted by
it is completely dominated by its dipole moment, and that
higher order moments such as quadrupole are negligible.  If an
object in vacuum has dipole moment $z$-component varying with
time as $p_z = p_{oz} \cos(\omega t)$ then the radiated power
per unit area of the detector is:\cite{griffiths99}
\begin{equation}
\label{1b.1}
S = \left(\frac{\mu_{o} p_{oz}^2 \omega^4}{32 \pi^2 c}\right) \frac{\sin^2 \theta}{r_{pd}^2}
\end{equation}
where $\mu_o$ and $c$ are the permeability and speed of light,
respectively, of free space.
$r_{pd}$ is the distance from the NP to the photon detector.
$\theta$ is the angle between the axis of the dipole and
the ray from the dipole to the photon detector.

Even though the results of this paper may be applied to the
situation of a NP in vacuum, it is more common in experimental
situations that the NP is embedded in a macroscopic matrix
such as a block of glass.  It is straightforward to adapt
Eq.~(\ref{1b.1}) to this situation:
\begin{equation}
\label{1b.2}
S = \left(\frac{\mu_{m} p_{oz}^2 \omega^4}{32 \pi^2 c_{m}}\right) 
\frac{\sin^2 \theta}{r_{pd}^2}
\end{equation}
where $\mu_{m}$ is the permeability of the matrix 
and $c_{m}$ is the speed of light in the matrix.
$c_{m} = 1/\sqrt{\mu_m \epsilon_m}$
and $\epsilon_m = \epsilon_{mro} \epsilon_o$,
where $\epsilon_{mro}$ is the relative permittivity of the matrix.
$\epsilon_o$ is the permittivity of free space.

As an aside, we add the ``o" to $\epsilon_{mro}$ to signify that
this is the equilibrium value of this quantity, which is
necessary because our later calculations can involve situations
where the vibrations of the NP cause variations in the density of
the matrix material and consequently result in changes of the
permittivity of the matrix in the immediate vicinity of the NP.
The formalism in later sections allows the permittivity of the
matrix in the immediate vicinity of the NP to be a function of
position.

When dealing with the macroscopic behavior of
dielectrics, charge may be viewed as ``free", ``bound", or
``total".  Charge which is artificially added to a preexisting
neutral dielectric material is certainly ``free".  The response
of the dielectric is to rearrange its own charge so as to
partly screen the ``free" charge.  Concentrations of this
redistributed charge are ``bound" charge.  Specifically, if
$\rho_f$, $\rho_b$, and $\rho$ are, respectively, the
free, bound, and total charge densities, then
$\rho_f + \rho_b = \rho$, $\nabla \cdot {\bf E} = \rho/\epsilon_o$
and $\nabla \cdot {\bf D} = \rho_f$.

A point charge $q_{f}$ (``f" stands for ``free") in vacuum
creates an electric field with magnitude $E = k q_f / r^2$
where $k = 1/(4 \pi \epsilon_o)$.
If the same point charge $q_{f}$ is artificially added into an
initially neutral block of dielectric material of relative
permittivity $\epsilon_{mro}$, it creates an electric field
$E = k q_f / (\epsilon_{mro} r^2)$.  The reduction in $E$ is due
to bound charge $q_b = -((\epsilon_{mro}-1)/\epsilon_{mro}) q_f$.
The total charge contained in the vicinity immediately
surrounding the free charge is $q = q_f + q_b$, and is given by
$q = q_f / \epsilon_{mro}$.

In like manner, when speaking of a point dipole in a
dielectric matrix, we must distinguish between the free,
bound, and total dipole moments.  Let these three (vector)
quantities be denoted respectively by ${\bf p}_f$, ${\bf p}_b$,
and ${\bf p}$.  For specificity, suppose that the dipole is
oriented along the $z$-axis with respective moments $p_{fz}$,
$p_{bz}$, and $p_{z}$.  If a free dipole $p_{fz}$ is
artificially inserted into a neutral dielectric, it will be
screened by bound dipole moment
$p_{bz} = -((\epsilon_{mro}-1)/\epsilon_{mro}) p_{fz}$.

In subsequent sections of this paper, we will always be
dealing with situations where there is no free charge
present.  As a result, all of the dipole moments which we
will calculate involve bound charge only.  When these
dipoles oscillate, they radiate electromagnetic energy.
This is the fundamental mechanism for all Rayleigh and
Raman scattering from NPs.  It is important to note that
Eq.~(\ref{1b.1}) and Eq.~(\ref{1b.2}) cannot be used to
find the energy radiated from such NPs because the
relative permittivity of the dielectric surrounding the
NP must be considered.

The correct way to obtain the radiated energy from a NP
embedded in a dielectric matrix is as follows.  Given the bound
dipole moment $p_{bz} = p_{z}$, we find the equivalent free
dipole ($p_{fz}$) which would create the same fields in the
immediate vicinity of the NP.  These are related by
$p_{fz} = \epsilon_{mro} p_z$.

Thus, in the absence of any free dipole moment, for an
oscillating point dipole $p_z = p_{oz} \cos(\omega t)$ in a 
dielectric matrix, the radiated power density is:
\begin{equation}
\label{1b.3}
S = \left(\frac{\mu_{m} (\epsilon_{mro}p_{oz})^2 \omega^4}{32 \pi^2 c_{m}}\right) \frac{\sin^2\theta}{r_{pd}^2}
\end{equation}

\section{Spherical Harmonic Transform}
\label{secII}

Consider a NP that is approximately spherical in shape and
approximately centered at the origin of the coordinate system.
The electric field is ${\bf E}({\bf r})$.  Because of the static
approximation, ${\bf E}=-\nabla V$ where $V({\bf r})$ is the
electric potential.

The motivation for introducing the word ``approximately'' twice
above is to justify the use of a spherical harmonic expansion
for the potential $V$ which is dominated by components with slow
angular variation.  In such a situation, we can hope for a
useful approximation with a finite number of spherical harmonics.
However, with generality any potential whatsoever could be
expressed in the form:
\begin{equation}
\label{Valm}
V(r,\theta,\phi)=
\sum_{\ell=0}^{\infty}
\sum_{m=-\ell}^{+\ell}
a_{\ell m}(r)S_{\ell m}(\theta,\phi)
\end{equation}
where $S_{\ell m}(\theta,\phi)$ are ``real spherical harmonics"
which are real-valued functions here defined as:
\begin{eqnarray}
S_{\ell m}=& -\sqrt{2}\,\mathrm{Re}(Y^m_{\ell}) & [m>0]  \nonumber \\
S_{\ell 0}=&Y_{\ell}^{0}                        & [m=0]  \nonumber \\
S_{\ell m}=& -\sqrt{2}\,\mathrm{Im}(Y^m_{\ell}) & [m<0]  \nonumber 
\end{eqnarray}

The $Y_{\ell}^m(\theta,\phi)$ are
conventionally defined complex valued spherical harmonic
functions.\cite{jackson85}

There are orthonormality conditions
\begin{equation}
\label{2.7}
\int \int S_{\ell m}S_{LM}\sin\theta d\theta d\phi=
\delta_{\ell L} \delta_{mM}
\end{equation}
where $\delta_{i j}$ is the Kronecker delta.
In particular, $S_{0\,0}\simeq 0.2821$,
$S_{1\,0}\simeq0.4886\,\cos\theta$,
$S_{1\,1}\simeq0.4886\,\sin\theta \cos\phi$, and
$S_{1\,\,\textrm{-1}}\simeq0.4886\,\sin\theta \sin\phi$.
Finally, note that:
\begin{equation}
\label{delSlm}
\nabla^2 S_{\ell m} = - \frac{\ell(\ell+1)}{r^2} S_{\ell m}
\end{equation}

In macroscopic electrostatics the permittivity $\epsilon$ and
the fields ${\bf E}$, ${\bf D}$ and the polarization ${\bf P}$
are related through ${\bf D} = \epsilon {\bf E}$
and ${\bf D} =\epsilon_o {\bf E} + {\bf P}$ so that
${\bf P} =\epsilon_o (\epsilon_r-1) {\bf E}$
where $\epsilon_o$ is the permitivity of free
space and $\epsilon=\epsilon_r \epsilon_o$, where $\epsilon_r$ is the
``relative permittivity."

The permittivity is a function of position, varying for two
reasons:
(1) The permittivity of the NP is different from the
permittivity of the surrounding glass matrix.
(2) Vibrations of the NP will lead to elastic strains
which will cause small time dependent variations of the
permittivity. (These periodic variations have a time scale on
the order of 3~ps, which is 1000 times longer than the period
of the incident laser light beam)

Motivated by the roughly spherical shape of the NP
about the origin, a real spherical harmonic expansion is
employed:
\begin{equation}
\label{2.10}
\epsilon(r,\theta,\phi)=\epsilon_o \sum q_{\ell m}(r)
S_{\ell m}(\theta,\phi)
\end{equation}

It also turns out to be convenient to introduce the logarithm
(base e) of the relative permittivity
$b=\log(\epsilon/\epsilon_o)$=$\log(\epsilon_r)$
which has the expansion:
\begin{eqnarray}
\label{2.11}
\log \left( \epsilon_r (r,\theta,\phi) \right) & = & b(r,\theta,\phi) \\
& = & \sum c_{\ell m} (r) S_{\ell m} (\theta,\phi) \nonumber
\end{eqnarray}

The permittivity can, in general be complex-valued, so
the complex analytic continuation of the logarithm function
will be used.
Note that the coefficients $c_{\ell m}(r)$ and $q_{\ell m}(r)$ are
dimensionless, while the $a_{\ell m}(r)$ are in volts.

The permittivity is assumed to be initially specified at all
points within the NP and the matrix.  The coefficients are
found using:
\begin{equation}
\label{qlmfromeps}
q_{LM}(r) = \frac{1}{\epsilon_o} \int \epsilon(r,\theta,\phi)
S_{LM} \sin\theta d\theta d\phi
\end{equation}
and
\begin{equation}
\label{clmfromepsr}
c_{LM}(r) = \int \log(\epsilon_r(r,\theta,\phi))
S_{LM} \sin\theta d\theta d\phi
\end{equation}

While $V({\bf r})$ may have non-analytic variation in the
vicinity of the surface of the NP due to rapid spatial
changes in $\epsilon({\bf r})$, we suppose that $V({\bf r})$
is smooth at the origin with the following series expansion:

\begin {equation}
\label{2.12}
\lim_{r \rightarrow 0}V(r,\theta,\phi)=\sum d_{\ell m} r^{\ell}
S_{\ell m}(\theta,\phi)
\end {equation}

Finally, we suppose that sufficiently far away from the
NP the permittivity again becomes constant.  In this
region the electric field is supposed to reach a spatially
constant value, but formally the potential can be expanded in
the large $r$ limit as:
\begin{equation}
\label{2.13}
\lim_{r \rightarrow \infty} V(r,\theta,\phi)=\sum e_{\ell m} r^{\ell}
S_{\ell m} (\theta,\phi)
\end{equation}

The $x$, $y$, and $z$ incident (far away) electric field
components are related to $e_{1 \, 0}$, $e_{1 \, 1}$, and
$e_{1 \, -1}$ as follows:
$E_{incz}\simeq-0.4886\,e_{1\,0}$,
$E_{incx}\simeq-0.4886\,e_{1\,1}$, and
$E_{incy}\simeq-0.4886\,e_{1\,\,\textrm{-1}}$.

\section{Integrating the Coulomb Equation}
\label{secIII}

The basis for this paper's calculations is that there are no
free charges associated with the NP, although there will be some
bound charge as a result of the induced polarization.  Thus
$\rho_f=0$, in which case $\nabla \cdot {\bf D}=0$ but
${\bf D}=\epsilon {\bf E}$, so
$\nabla \cdot (\epsilon {\bf E})=0$. Note further that
${\bf E}=-\nabla V$, so that
$\nabla \cdot(\epsilon(\nabla V))=0$.
Differentiating,
\begin{equation}
\epsilon \nabla^2 V + (\nabla V)\cdot(\nabla \epsilon)=0
\end{equation}

Next, divide this through by $\epsilon({\bf r})$.
If $b=\log(\epsilon/\epsilon_o)=\log(\epsilon_r)$
then $\nabla b = (\nabla \epsilon)/\epsilon$.
In this case
\begin{equation}
\label{3.5}
\nabla^2 V + (\nabla V)\cdot(\nabla b)=0
\end{equation}

The boundary value problem that we are solving is as follows:
Supposing $b({\bf r})$ to be known, it will be Eq.~(\ref{3.5})
which will be solved to yield $V({\bf r})$.  The boundary
condition is that ${\bf E}$ approaches a specified constant
value ${\bf E}_{inc}$ in the limit of large $r$.

To do this, we next insert the real spherical harmonic
expansions of $V$ and $b$ into Eq.~(\ref{3.5}).  The result
is multiplied by a given $S_{\ell m}(\theta,\phi)$, and then
integrated over $\theta$ and $\phi$.  This yields a set
(indexed by $\ell$ and $m$) of coupled
ordinary differential equations as follows:
\begin{multline}
\label{3.6}
-r^2 a''_{\ell m} = 2ra'_{\ell m}-\ell(\ell+1)a_{\ell m} \\
+
\sum_{LM} \sum_{\lambda \mu}
\biggl\{ a'_{LM}c'_{\lambda \mu}H(\ell m;LM;\lambda \mu)   \\
+ a_{LM}c_{\lambda \mu} K(\ell m|LM;\lambda \mu) \biggr\}
\end{multline}
where
\begin{equation}
\label{defineH}
H(\ell m; LM; \lambda \mu)=\int S_{\ell m} S_{LM} S_{\lambda \mu} d \Omega
\end{equation}
and
\begin{multline}
\label{defineK}
K(\ell m | LM; \lambda \mu) = 
\int S_{\ell m} \left(
\frac{\partial}{\partial \theta} S_{LM}
\frac{\partial}{\partial\theta} S_{\lambda \mu} \right. \\
+ \left. \frac{1}{\sin^2\theta}
\frac{\partial}{\partial\phi} S_{LM} \frac{\partial}{\partial\phi}
S_{\lambda \mu} \right) d \Omega
\end{multline}
where $d \Omega$ denotes $\sin\theta d\theta d\phi$.
The constants $H(;;)$ and $K(| ;)$ have to be numerically
integrated in advance and stored in a lookup table.
$a''_{\ell m}$ is the second derivative and $a'_{\ell m}$
is the first derivative of $a_{\ell m}(r)$.

For the initial small $r$ value in each integration of
Eq.~(\ref{3.6}), only one of the constants $d_{\ell m}$
(from Eq.~(\ref{2.12}))
will be 1, and the others are all zero.  (Later on, we will be able to
determine the actual values of all of the $d_{\ell m}$ so as to
satisfy the large-$r$ boundary conditions of the original
boundary value problem.)

For a given choice of initial conditions in terms of the
constants $d_{\ell m}$, these coupled second order ordinary
differential equations, (Eq.~(\ref{3.6})), can be integrated
outwards from $r$=0 until large $r$ is reached, at which point
the constants $e_{\ell m}$ can be found.

Because of the linearity of the equations, these are related by
\begin{equation}
\label{3.10}
e_{\ell m}=\sum \sum F_{\ell m L M} d_{LM}
\end{equation}
where $F$ is a matrix.  All of the coefficients of $F$ can be
calculated through successive integrations of the coupled
equations (i.e. varying $\ell$ and $m$).

Suppose the desired solution corresponds to the case where
$e_{1\,0}\simeq-2.047\,E_{incz}$ and all other $e_{\ell m}$
are 0. Let $G$ be the inverse of the matrix $F$. Then
\begin{equation}
d_{\ell m}=\sum \sum G_{\ell m L M} e_{LM}
\end{equation}

The next step is to repeat the integration of the coupled
differential equations with the resulting values of
$d_{\ell m}$, in which case all of the values of
$a_{\ell m}(r)$ can be found for all values of $r$.
Using Eq.~(\ref{Valm}), this then yields $V({\bf r})$, from which
the electric field, polarization, bound charge density and
dipole moment can all be calculated.

\section{Objects in Vacuo}
\label{secV}

In order to understand how the dipole moment of a NP
in a dielectric is calculated later in this paper,
we first review the simpler case of a NP in a vacuum.
The dipole moment of the NP is ${\bf p}$, with
Cartesian components $p_x$, $p_y$, and $p_z$.  Since there is no
free charge, the dipole moment arises from the bound charge,
whose volume density is $\rho_b({\bf r})$.
The polarization of the material is ${\bf P}({\bf r})$, and
$\rho_b=- \nabla \cdot {\bf P}$.  Since polarization is dipole
moment per unit volume, dipole moment can also be calculated
from:
\begin{equation}
\label{pfromP}
{\bf p}= \int {\bf P} d^3 r
\end{equation}
But ${\bf p}$ can also be obtained directly from its definition,
given that $\rho = \rho_b$ in this situation:
\begin{equation}
\label{pfromrho}
{\bf p}=\int {\bf r} \rho_b \, d^3 r
\end{equation}

In addition, since $\nabla^2 V = -\rho / \epsilon_o$
\begin{equation}
\label{pfromV}
{\bf p}= - \epsilon_o \int {\bf r} \nabla^2 V d^3 r
\end{equation}

Equations~(\ref{pfromP}), (\ref{pfromrho}), and (\ref{pfromV})
will serve as the basis of three independent numerical methods
for calculating the dipole moment (and hence the polarizability
tensor) of a NP.  These methods are presented in
Sections \ref{secX}, \ref{secXI}, and \ref{secXII}, respectively.
This threefold redundancy of our method serves
as a check on the numerical reliability of its results.

To see that Eq.~(\ref{pfromP}) and Eq.~(\ref{pfromrho}) give the
same answer, consider an arbitrary
closed region $U$ enclosing the NP in its interior.  Let
$\partial U$ denote the surface of $U$.  In this case, ${\bf P}$
is zero everywhere on $\partial U$.

Next, consider the vector field ${\bf B}=z{\bf P}$, where $z$ is
the $z$-component of position. According to the divergence theorem:
\begin{equation}
\label{5.2b}
\int_{\partial U} {\bf B} \cdot {\bf dA}= \int_U (\nabla \cdot {\bf B}) d^3 r
\end{equation}
Therefore, $\int (\nabla \cdot {\bf B}) d^3 r=0$,
since ${\bf P}={\bf 0}$ on the boundary of this closed region.

However, since
$\nabla \cdot (z {\bf P}) =P_z + z \nabla \cdot {\bf P}$
thus
\begin{equation}
\label{5.6}
{\bf p}= \int_U {\bf r} \rho_b d^3 r = \int_U {\bf P} d^3 r
\end{equation}

\section{Embedded Dipoles}
\label{secVI}

Unfortunately, the formulas of Section~\ref{secV} do not
all apply to the situation of interest for an embedded NP,
where the surrounding material (a glass matrix, for example)
has a susceptibility, so that the polarization in the matrix is
nonzero, and must be taken into account when finding the dipole
moment of the NP.  The dipole moment that is relevant to light
scattering experiments on NPs is the additional dipole moment
produced as a result of the presence of the NP.  The
relative permittivity of the glass matrix without density
variations is
$\epsilon_{mro}$.  The polarization of the glass
matrix in the absence of the NP would be as follows:
${\bf P}_{inc}=\epsilon_o (\epsilon_{mro} - 1) {\bf E}_{inc}$
This is uniform, so $\nabla \cdot {\bf P}_{inc}= 0$ and
consequently the formula (Eq.~(\ref{pfromrho}))
${\bf p}= \int {\bf r} \rho_b \, d^3 r$
can still be used in this more general case.

Because ${\bf P}$ is nonzero throughout the matrix,
Eq.~(\ref{pfromP}) cannot be used for a NP embedded in an
infinite matrix (since the integral would diverge).
Here, we derive a formula in terms of ${\bf P}$ that can
be used to find the dipole moment in this case.

Once again, let us define the vector field
${\bf B}$ to be ${\bf B}=z{\bf P}$. Then choose a spherical
region $U$ of radius $R$ centered on the NP, where $R$
is large enough that density fluctuations due to vibrations of
the NP are negligible, and also large enough that
fields due to multipole moments beyond dipole order can be
ignored.
Again, we can use the divergence theorem, Eq.~(\ref{5.2b}).

First, we evaluate the left side of Eq.~(\ref{5.2b}).  We need
to know {\bf B} on the spherical surface $\partial U$
that surrounds the NP. The NP has dipole moment $z$-component
$p_z$.  The potential due to this dipole (not including that
resulting from the incident field) is:\cite{griffiths99}
\begin{equation}
\label{6.4}
V_{dip}= \frac{1}{4 \pi \epsilon_o} \frac{p_z \cos\theta}{r^2}
\end{equation}

The subscript `\textit{dip}' stands for `dipole'.
[Note: It may be tempting to insert a factor
of $\epsilon_{mro}$ into the denominator here. But remember
that this is not the potential due to free charge embedded in
a dielectric. There is no free charge. To find the potential
due to a dipole, we can apply the formula
$\nabla^2 V= -\rho / \epsilon_o$
where $\rho$ is the charge density, the sum of free and bound charge.]
The electric field created in the matrix by the dipole moment
of the NP has radial component
$E_{dipr}=-\partial V_{dip} / \partial r$
\begin{equation}
\label{6.5}
E_{dipr}= \frac{1}{4 \pi \epsilon_o} \frac{2 p_z \cos\theta}{r^3}
\end{equation}

The part of the polarization in the matrix created by the
NP has radial component
\begin{equation}
\label{6.6}
P_{dipr}= \frac{(\epsilon_{mro}-1)}{4 \pi}
\frac{2 p_z \cos\theta}{r^3}
\end{equation}

The vector field ${\bf B}$ is the sum of the part created by the 
NP and the part due to the incident applied field:
\begin{equation}
\label{6.7}
{\bf B} ={\bf B}_{dip} + {\bf B}_{inc} 
\end{equation}

The radial component of ${\bf B}_{dip}$ is:
\begin{equation}
\label{6.8}
B_{dipr}= \frac{(\epsilon_{mro}-1)}{4 \pi}
\frac{2 p_z \cos^2\theta}{r^2}
\end{equation}

${\bf B}_{dip} \cdot d {\bf A}$ is equal to $B_{dipr} dA$.
We now want to integrate this over the sphere of radius $R$.
\begin{align}
\notag
\int {\bf B}_{dip} \cdot d {\bf A} &= 
\int B_{dipr} R^2 \sin\theta d \theta d \phi \\
\notag &= \int \frac{ (\epsilon_{mro}-1)}{4 \pi}
\frac{2 p_z \cos^2\theta}{R^2}
R^2 \sin\theta d \theta d \phi \\
\notag &= \int(\epsilon_{mro}-1) p_z \cos^2\theta \sin\theta d \theta \\
\label{6.9}  &= \frac{2(\epsilon_{mro}-1)}{3}p_z 
\end{align}

As for the incident field part ${\bf B}_{inc}$, it is:
\begin{equation}
\label{6.10}
{\bf B}_{inc} =z {\bf P}_{inc}= z \epsilon_o (\epsilon_{mro} - 1)
{\bf E}_{inc}
\end{equation}

To evaluate the surface integral
$\int {\bf B}_{inc} \cdot {\bf dA}$
we need only the radial component:
\begin{equation}
\label{6.11}
B_{incr}=\epsilon_o (\epsilon_{mro} - 1) r \cos\theta E_{incz} \cos\theta
\end{equation}
where $E_{incz}$ is the $z$-component of the incident field. We
need to integrate this over the same sphere of radius $R$:
\begin{align}
\notag
\int {\bf B}_{inc} \cdot d {\bf A}
&= \epsilon_o (\epsilon_{mro}-1) \int E_{incz}
R^3 \cos^2\theta \sin\theta d\theta d\phi \\
\notag &= \frac{4\pi}{3} (R^3 \epsilon_o (\epsilon_{mro}-1) E_{incz}) \\
\label{6.12} &= \int P_{incz} d^3 r 
\end{align}
where $P_{incz}=\epsilon_o (\epsilon_{mro} - 1) E_{incz}$ is the
polarization in the matrix if the NP were not present.

We now combine results:
\begin{align}
\notag
\int {\bf B} \cdot d {\bf A}
&= \int {\bf B}_{dip} \cdot d {\bf A} + \int {\bf B}_{inc} \cdot d {\bf A} \\
\label{6.13} &= \frac{2(\epsilon_{mro} -1)}{3} p_z + \int P_{incz} d^3 r 
\end{align}

Once again, because 
$\nabla \cdot (z {\bf P})= P_z + z \nabla \cdot {\bf P}$
then
\begin{align}
\notag
\int {\bf B} \cdot d {\bf A} &= \int (\nabla \cdot {\bf B} ) d^3 r\\
\label{6.15} &= \int z (\nabla \cdot {\bf P} ) d^3 r + \int P_z d^3 r
\end{align}

Next, since $\rho_b = - \nabla \cdot {\bf P}$ and
$p_z= \int z \rho_b d^3 r$, we get
\begin{equation}
\label{6.16}
\frac{2(\epsilon_{mro}-1)}{3} p_z + \int P_{incz}d^3 r=-p_z + \int P_z d^3 r
\end{equation}
and finally
\begin{equation}
\label{6.17}
p_z = \frac{3}{2 \epsilon_{mro}+ 1} \int (P_z- P_{incz}) d^3 r
\end{equation}
which can then be vector generalized as follows:
\begin{equation}
\label{6.18}
{\bf p} = \frac{3}{2 \epsilon_{mro}+ 1}
{\bf \int} ({\bf P}- \epsilon_o (\epsilon_{mro}-1){\bf E}_{inc} ) d^3 r
\end{equation}

\section{Dipole Moment: Polarization}
\label{secX}

Finally we address the central problem of calculating the dipole
moment of an object with inhomogeneous permittivity.  The object
has an arbitrary shape. In the case of a vibrating NP, the
region of inhomogeneous permittivity extends outside the NP into
the glass matrix since density variations resulting from the
vibrations will affect the permittivity of the glass as well as
that of the NP.  Only at sufficient distance from the NP will
the relative permittivity reach its homogeneous value of
$\epsilon_{mro}$.  Note that $\epsilon_{mro}$ is the square of
the index of refraction of the glass.

There is some given incident electric field ${\bf E}_{inc}$.
By repeating calculations of ${\bf p}$ for different
orientations of ${\bf E}_{inc}$ we can obtain the polarizability
tensor $\alpha$.

We begin by calculating $p_z$ using Eq.~(\ref{6.17}).
We first make the substitution $E_z = - \partial V / \partial z$,
yielding
\begin{equation}
\label{10.2}
p_z= \frac{3\epsilon_o}{2 \epsilon_{mro}+1} \int \biggl\{ (1 - \epsilon_r)
\frac{\partial V}{\partial z} + (1- \epsilon_{mro}) E_{incz} \biggr\} d^3 r
\end{equation}

The following seven steps will re-express
Eq.~(\ref{10.2}) in terms of an explicit quadruple summation
over indices $\ell$, $m$, $L$, and $M$, as shown in Eq.~(\ref{10.13}):

Step 1: Recall the expansion of $\epsilon_r$ in terms of
the summation of the $q_{\ell m}$ in Eq.~(\ref{2.10}). 

Step 2: We want to put the summation for ($1-\epsilon_r$)
into a neat form.  Note that $1=\sqrt{4\pi}S_{00}$, so that
\begin{equation}
\label{oneminusepsr}
(1-\epsilon_r)=
\sum_{\ell=0}^{\infty} \sum_{m=-\ell}^{+\ell}
[\sqrt{4 \pi} \delta_{\ell 0} - q_{\ell m}] S_{\ell m}
\end{equation}

Step 3: Note that
\begin{equation}
\frac{\partial V}{\partial z}= \cos\theta \frac{\partial V}{\partial r}-\frac{\sin\theta}{r} \frac{\partial V}{\partial \theta}
\end{equation}

Step 4: Recall the expansion of $V$ in terms of the
summation of the $a_{\ell m}$ in Eq.~(\ref{Valm}).

Step 5: Take the derivative with respect to $r$:
\begin{equation}
\label{10.11}
\frac{\partial V}{\partial r} = \sum_{L=0}^{\infty} \sum_M a'_{LM} (r) S_{LM}(\theta, 
\phi)
\end{equation}

Step 6: And with respect to $\theta$ as well:
\begin{equation}
\label{10.12}
\frac{\partial V}{\partial \theta} = \sum_{L=0}^{\infty} \sum_M a_{LM} (r)
\left( \frac{\partial S_{LM}}{\partial \theta} \right)
\end{equation}

Step 7: Substitute Eq's.~(\ref{oneminusepsr}), (\ref{10.11}), and
(\ref{10.12}) all into Eq.~(\ref{10.2}) so as to get:
\begin{multline}
\label{10.13}
p_z=\frac{3\epsilon_o }{2 \epsilon_{mro}+1} \int \sum_{\ell m L M} \biggl\{ 
[\sqrt{4 \pi} \delta_{\ell 0}  \\
- q_{\ell m}(r)] S_{\ell m} (\theta, \phi)
\cos\theta a'_{LM}(r) S_{LM}(\theta, \phi) \\
- [\sqrt{4 \pi} \delta_{\ell 0} - q_{\ell m}(r)] S_{\ell m} (\theta, \phi)
\frac{\sin\theta}{r} a_{LM}(r) \frac{\partial S_{LM}}{\partial \theta}  \\
+ \delta_{\ell 0} \: \delta_{m 0} \: \delta_{L 1} \: \delta_{M 0}
(1-\epsilon_{mro})E_{incz} \biggr\} r^2 dr d\Omega
\end{multline}
where $d^3 r$ has been replaced with $r^2dr d\Omega$ and
$d\Omega= \sin\theta d\theta d\phi$.  Next, we want to carry
out the angular part of the integration.  First, note that
$\cos\theta\simeq 2.047\,S_{1\,0}=\sqrt{4 \pi/3} S_{1\,0}$
and second note that
$\sin\theta=-(d/d\theta)\cos\theta=-\sqrt{4 \pi/3}
(\partial/\partial\theta) S_{1\,0}$.
Third, refer to the definitions of $H(;;)$ in
Eq.~(\ref{defineH}) and $K(|;)$ in Eq.~(\ref{defineK}).
Putting these all together,
\begin{multline}
\label{10.14}
p_z=
\frac{3\epsilon_o }{2 \epsilon_{mro}+1}
\int \sum_{\ell m} \sum_{LM}
\biggl\{
\sqrt{\frac{4 \pi}{3}}[\sqrt{4 \pi} \delta_{\ell 0} \\
- q_{\ell m}] \Bigl( a'_{LM} H(LM;1 0;\ell m)  \\
+ \frac1r a_{LM} K(\ell m|1 0;LM) \Bigr)   \\
+ 4 \pi \: \delta_{\ell 0} \: \delta_{m 0} \: \delta_{L 1} \: \delta_{M 0}
(1-\epsilon_{mro}) E_{incz} \biggr\} r^2 dr
\end{multline}

Note that Eq.~(\ref{10.14}) still applies in the situation
where $E_{incz}$=0 and $E_{incx}$ or $E_{incy}$ is nonzero.
This would be the situation when calculating the off-diagonal
elements of the polarizability, $\alpha$.

The other components $p_x$ and $p_y$ can be found by analogous
expressions: $x \Leftrightarrow 1$ and $y \Leftrightarrow -1$.
\begin{multline}
\label{10.15}
p_x=
\frac{3\epsilon_o }{2 \epsilon_{mro}+1}
\int \sum_{\ell m} \sum_{LM}
\biggl\{
\sqrt{\frac{4 \pi}{3}}[\sqrt{4 \pi} \delta_{\ell 0}  \\ 
- q_{\ell m}] \Bigl( a'_{LM} H(LM;1 1;\ell m)  \\
+ \frac{1}{r} a_{LM} K(\ell m|11;LM) \Bigr)  \\
+ 4 \pi \: \delta_{\ell 0} \: \delta_{m 0} \: \delta_{L 1} \: \delta_{M 0}
(1-\epsilon_{mro}) E_{incx} \biggr\} r^2 dr
\end{multline}
\begin{multline}
\label{10.16}
p_y=
\frac{3\epsilon_o }{2 \epsilon_{mro}+1}
\int \sum_{\ell m} \sum_{LM}
\biggl\{
\sqrt{\frac{4 \pi}{3}}[\sqrt{4 \pi} \delta_{\ell 0}  \\ 
- q_{\ell m}] \Bigl( a'_{LM} H(LM;1 \,\textrm{-1};\ell m) \\
+ \frac{1}{r} a_{LM} K(\ell m|1 \,\textrm{-1};LM) \Bigr)  \\
+ 4 \pi \: \delta_{\ell 0} \: \delta_{m 0} \: \delta_{L 1} \: \delta_{M 0}
(1-\epsilon_{mro}) E_{incy} \biggr\} r^2 dr
\end{multline}

The dipole moment ${\bf p}$ and applied electric field 
${\bf E}_{inc}$ are related by ${\bf p}=\alpha {\bf E}_{inc}$ 
where $\alpha$ is the polarizability tensor of the NP. 

It makes things more compact to label vector components by the
subscript $\mu$ which runs from $-1$ to 1. For example, ${\bf p}$
has components denoted generally by $p_{\mu}$ and specifically
$p_{\textrm{-1}}$, $p_0$ and $p_1$, where the correspondence to usual
Cartesian notation is as follows:
$(p_{1},p_{\textrm{-1}},p_{0})$=$(p_{x},p_{y},p_{z})$ and
$(E_{inc1},E_{inc\textrm{-1}},E_{inc0})$=$(E_{incx},E_{incy},E_{incz})$

In this way, it is possible to write a single formula which can
calculate any of the three components of the dipole moment of
the NP, where $\mu \in \{ 1, -1, 0\}$:
\begin{multline}
\label{10.20}
p_{\mu}=
\frac{3\epsilon_o }{2 \epsilon_{mro}+1}
\int
\biggl( \sum_{\ell m} \sum_{LM} \biggl\{
\sqrt{\frac{4 \pi}{3}}[\sqrt{4 \pi} \delta_{\ell 0}  \\
- q_{\ell m}(r)]\Bigl( a'_{LM}(r) H(LM;1 \mu;\ell m) \\
+ \frac{1}{r} a_{LM}(r) K(\ell m|1\, \mu;LM) \Bigr)\biggr\}  \\
+ 4 \pi (1-\epsilon_{mro}) E_{inc \mu} \biggr) r^2 dr
\end{multline}

\section{Dipole Moment: Charge}
\label{secXI}

For redundancy, the calculation of Section~\ref{secX} is
repeated, but this time based on the charge-based expression for
dipole moment, Eq.~(\ref{pfromrho}).  The bound charge density is
$\rho_b= -\nabla \cdot {\bf P}$ where the polarization ${\bf P}$
is given by ${\bf P}=\epsilon_o (\epsilon_r-1) {\bf E}$.  Also,
${\bf E}= - \nabla V$. So
\begin{equation}
\label{11.1}
\rho_b=\epsilon_o \nabla \cdot((\epsilon_r - 1) \nabla V)
\end{equation}
\begin{equation}
\label{11.2}
\rho_b=\epsilon_o \{ (\epsilon_r-1) \nabla^2 V+ (\nabla V) \cdot (\nabla 
\epsilon_r) \}
\end{equation}

From this, and using Eqs.~(\ref{Valm}) and (\ref{oneminusepsr}),
\begin{multline}
\label{11.3}
\rho_b=\epsilon_o
\biggl\{
\sum \sum ( q_{\ell m} - \sqrt{4 \pi} \delta_{\ell m}) S_{\ell m} \nabla^2 (a_{LM} S_{LM}) \\
+ \sum \sum \frac{\partial}{\partial r} (q_{\ell m} S_{\ell m})
\frac{\partial}{\partial r} (a_{LM} S_{LM})  \\
+\sum \sum \frac{1}{r^2} \frac{\partial}{\partial \theta} (q_{\ell m}
S_{\ell m}) \frac{\partial}{\partial \theta} (a_{LM} S_{LM}) \\
+ \sum \sum \frac{1}{r^2 \sin^2\theta}
\frac{\partial}{\partial \phi}
(q_{\ell m} S_{\ell m}) \frac{\partial}{\partial \phi} (a_{LM} S_{LM})
\biggr\}
\end{multline}

Carrying out these derivatives,
and also using Eq.~(\ref{delSlm}), we get
\begin{multline}
\label{11.8}
\rho_b= 
\epsilon_o \biggl\{
\sum \sum (q_{\ell m}-\sqrt{4 \pi} \delta_{\ell 0} ) \frac{1}{r^2}
[r^2 a''_{LM}+2r a'_{LM}  \\
- L(L+1)a_{LM}] S_{\ell m} S_{LM} \\
+ \sum \sum q'_{\ell m} a'_{LM} S_{\ell m} S_{LM} \\
+ \sum \sum q_{\ell m} a_{LM} (\frac{1}{r^2}) \left[ 
\frac{\partial S_{\ell m}}
{\partial \theta} \frac{\partial S_{LM}}{\partial \theta}  \right. \\
\left. + \frac{1}{\sin^2\theta} \frac{\partial S_{\ell m}}{\partial \phi}
\frac{\partial S_{LM}}{\partial \phi}  \right]
\biggr\}
\end{multline}

This last equation can now be inserted into Eq.~(\ref{pfromrho}) to
obtain the dipole moment.

Suppose we want to calculate a single component of the dipole
moment, $p_\mu$ where $\mu$ could be -1, 0 or 1. If $\mu=0$ then
we multiply $\rho_b$ by $z=r \cos (\theta)\simeq2.047\,r S_{1\,0}$.
In general we multiply by $2.047\,r S_{1 \mu}$.  The
integrand is then integrated over all space.  It is convenient
to carry out the angular integrations first:
\begin{multline}
\label{11.10}
p_{\mu}=\sqrt{\frac{4 \pi}{3}} \epsilon_o \int \sum_{\ell m}
\sum_{L M} \biggl\{
(q_{\ell m}- \sqrt{4 \pi} \delta_{\ell 0}) [r^2 a''_{LM} 
+ 2ra'_{LM}  \\
-L(L+1)a_{LM}] H(1 \mu; L M;\ell m) \\
+ r^2 q'_{\ell m} a'_{LM} H(1 \mu; L M; \ell m)  \\
+ q_{\ell m} a_{LM}K ( 1 \mu| L M;\ell m)
\biggr\} r dr
\end{multline}

Results obtained using this equation can be compared to those
obtained from Eq.~(\ref{10.20}).  This provides a guard against
programming errors and numerical problems in the integrations
over $r$.

\section{Dipole Moment: Potential}
\label{secXII}

A third independent method for obtaining dipole moment is
presented in this section.  Starting from Coulomb's law in
differential form:
$\nabla \cdot {\bf E}= \rho / \epsilon_o$, and using
${\bf p} =\int {\bf r} \rho d^3 r$ as well as ${\bf E}= -\nabla V$,
we get:
\begin{equation}
\label{12.1}
{\bf p} =- \epsilon_o \int {\bf r} (\nabla^2 V) r^2 dr d\Omega
\end{equation}

Next, use Eq.~(\ref{Valm}) to express $V$ in terms of $a_{\ell m}$
and real spherical harmonics. Also, specialize to $p_z$.
\begin{equation}
\label{12.2}
p_z=- \epsilon_o {\bf \int} \sum_{\ell m} r \cos\theta \nabla^2 [a_{\ell m}
S_{\ell m}] r^2 dr d\Omega
\end{equation}

Recall that $\cos\theta\simeq2.047\,S_{1\,0}$.  Also, make use
of Eq.~(\ref{delSlm}).  So this becomes:
\begin{multline}
\label{12.3}
p_z =
-2.047 \, \epsilon_o \int \sum_{\ell m} [\frac{1}{r^2}
\frac{d}{dr}(r^2 \frac{d}{dr} a_{\ell m})  \\
-\frac{\ell(\ell+1)}{r^2} a_{\ell m}] S_{1\,0} S_{\ell m} r^3 dr d\Omega
\end{multline}

It is now apparent how to generalize from the $z$-axis to the
$\mu$ axis where $\mu$ is -1, 0, or 1. Also, the angular
integration can be performed (employing Eq.~(\ref{2.7})),
followed by the summation over $\ell$ and $m$:
\begin{equation}
\label{12.5}
p_\mu=-\sqrt{\frac{4 \pi}{3}} \epsilon_o \int [r^3 a''_{1 \mu} +
2r^2 a'_{1 \mu} - 2 r a_{1 \mu}] dr
\end{equation}

Equation~(\ref{12.5}) for dipole moment is far simpler than
Eq.~(\ref{10.20}) and Eq.~(\ref{11.10}). In particular, the
$q_{\ell m}$ coefficients are no longer required. In theory the
range of integration is from $r=0$ to $\infty$.  However,
Eq.~(\ref{12.5}) is more susceptible to numerical error buildup
in the high $r$ region compared to the other two methods, and
the upper limit of integration has to be carefully restricted.
The integration step should also be kept small. Even so, it
tends to give much more accurate results than the other two
methods methods when tested on prolate ellipsoids.

\section{Numerical Methods}
\label{secXIII}

In practice, the expansion for $V$ in Eq.~(\ref{Valm}) has a
limited value of $\ell$, here denoted as $\ell_{maxa}$.  In a
similar way, the expansion for $\log (\epsilon_r)$ in
Eq.~(\ref{2.11}) is limited to a finite value of $\ell$,
denoted $\ell_{maxc}$.

It was found that $\ell_{maxc}$ must be chosen jointly with
$\ell_{maxa}$.  Best results are obtained when $\ell_{maxc}$ is
less than $\ell_{maxa}$.  Further increasing $\ell_{maxa}$
while holding $\ell_{maxc}$ constant does not improve the
results.

It is desirable to make careful comparisons with the exact
solution for a homogeneous dielectric ellipsoid in a homogeneous
matrix.  However, in such a situation the permittivity is a
singular function which is not suitable for integration.  To get
around this problem, the dielectric ellipsoid is approximated by
a smoothed permittivity.  The smoothing must be carried out over
a short distance (surface thickness) in order to accurately
approximate the ideal shape.  In addition, this smoothing should
be done in such a way that the derivatives of permittivity are
also smooth. This is implemented in a C++ program named
\textit{varpr27e}.  However when the eccentricity is nonzero the
functions become
smooth, and in these cases even zero-order smoothing (continuous,
but 1st derivative discontinuous) seems to work.

The coefficients $c_{\ell m}$ and $q_{\ell m}$ that are obtained
from the smoothed permittivity function are now smooth functions
of $r$. Even so, they change sharply near certain points.  It is
necessary to use integration with variable step size where a
much smaller step is used near the difficult points.

It is practical to first obtain the functions $c_{\ell m}$ and
$q_{\ell m}$.  These functions of $r$ can be stored in a look-up
table.
However, because of
the sharp variation at certain points the step size between
entries in the table must be variable. This is taken care
of in the program \textit{varpr27e} which generates a
file as output which is read by \textit{varpr28j}.

The lookup table for $H(;;)$ and $K(|;)$
can grow to several hundred entries, and access time 
can be minimized through use of a hash index (likely 
first guess) method as implemented in the functions  
$H(;;)$ and $K(|;)$ in \textit{varpr28j}.      
Most computation time ends up being spent in the evaluation of 
the right side of Eq.~(\ref{3.6}) because of the triple 
nested loop required. This is why the access time of $H(;;)$ and
$K(|;)$ is so important.

It is preferable to carry out the actual numerical integration
of Eq.~(\ref{3.6}) in terms of the variables $w_{\ell m}(r)$,
such that $a_{\ell m}(r)= r^{\ell} w_{\ell m}(r)$.  This is
because these variables are constant both at large and small $r$.

It is not possible to handle the situation where the
permittivity inside the NP is a purely negative number. That
is because smoothing at the boundary would result in a point
where the permittivity is zero, so that the logarithm is
singular.  However, if the permittivity inside is complex valued
then this problem is avoided.

\section{Checks of Correctness}

The three methods for determining the dipole moment
given as Eqs.~(\ref{10.20}), (\ref{11.10}), and (\ref{12.5}),
do not give identical results when finite $\ell$ cutoffs
( $\ell_{maxa}$ and $\ell_{maxc}$) are used.
However, mutual convergence of the three methods to calculate
dipole moment is an indication of convergence as the
two $\ell$ cutoffs are increased.

Given the length of this paper, we have chosen not to present
specific numerical results.  But we can mention the numerical
tests that we have done as a check of correctness of the
equations presented here.  We compared our numerical results
to exact ones for the (1) dielectric sphere, with
positive and complex valued permittivity (2) prolate spheroid
(3) oblate spheroid and (4) dielectric sphere with its center
not located at the origin of coordinates.

All of these cases have an exact solution that comes from
Eq.~(8.10) on page 41 of Landau's book.\cite{landau84}
This is for the dipole moment of a homogeneous ellipsoid
(with semiaxes $a$, $b$, and $c$)
in a vacuum with an asymptotically uniform electric field
along the $z$-axis.  When translated from cgs to SI units
and into our notation it becomes:
\begin{equation}
\label{14.1}
p_z=(volume) E_{incz} 
\frac{\epsilon_o (\epsilon_{inr}-1)}{1+(\epsilon_{inr}-1)n^{(z)}}
\end{equation}
where $n^{(z)}$ is the $z$-axis depolarization factor which is
equal to 1/3 for a sphere and ``\textit{volume}" is
$(4 \pi /3 ) abc$.  $\epsilon_{inr}$ is the relative
permittivity of the ellipsoid.  Using our Eq.~(\ref{8.7}) (in
the Appendix) the formula for the case of a homogeneous
dielectric ellipsoid embedded in a homogeneous dielectric is:   
\begin{equation}
\label{14.2}
p_z=(volume) E_{incz}
\frac{\epsilon_o (\epsilon_{inr}-\epsilon_{mro})}{\epsilon_{mro}+(\epsilon_{inr}-\epsilon_{mro})n^{(z)}}
\end{equation}

For an oblate spheroid where $a=b$ and $c<a$, the
eccentricity is $e=\sqrt{a^2/b^2-1}$.
\begin{equation}
\label{14.3}
n^{(z)}=\frac{1+ e^2}{e^3}(e- \arctan (e))
\end{equation}
\begin{equation}
\label{14.4}
n^{(x)}=n^{(y)}=\frac{1}{2}(1-n^{(z)})
\end{equation}

There is an increase in numerical error as the condition for
dipole surface plasmon resonance is approached.  This is because
$F_{1010}$ approaches zero after many positive and negative
contributions so that numerical errors become predominant
when $\ell$ cutoffs are used.

In order to see the pattern of convergence as $\ell_{maxc}$
and $\ell_{maxa}$ increase, some series of values have been
calculated to see if the percentage error goes to zero.  The
percentage error does decrease monotonically.

\textbf{Acknowledgments:} This work was supported by the Natural
Sciences and Engineering Research Council of Canada.

\appendix

\section{Homogeneous Objects}
\label{secVIII}

We present a fourth method to calculate the dipole moment,
however this one is specialized to the situation of a
homogeneous dielectric embedded in a homogeneous dielectric.
There is no bound charge density $\rho_b$ at interior points of
a homogeneous dielectric, but bound surface charge $\sigma_b$
can exist at dielectric surfaces. To see why, note that
${\bf P}=(\epsilon_r-1))/\epsilon_r) {\bf D}$, and
$\nabla \cdot {\bf D}=\rho_f$ where $\rho_f$ is the free charge
density.  In a region where $\epsilon_r$ is constant and there
is no free charge $\nabla \cdot {\bf D}=0$
so that $\nabla \cdot {\bf P}=0$ as well. Thus $\rho_b=0$.

In this case, the volume integral for dipole moment
${\bf p}= \int {\bf r} \rho_b d^3 r$ can be replaced with the
surface integral
\begin{equation}
\label{8.1}
{\bf p}= \int {\bf r} \sigma_b dA
\end{equation}

The bound surface charge density is given by
\begin{equation}
\label{8.2}
\sigma_b= P_{inn} - P_{outn}
\end{equation}
where $P_{outn}$ is the normal (outward pointing) component of
the polarization just outside the surface, while $P_{inn}$ is
the normal (outward pointing) component of the polarization just
inside the surface.
This can be related to the electric field at the surface since
${\bf P}=\epsilon_o (\epsilon_r - 1) {\bf E}$.  Thus:
\begin{equation}
\label{8.3}
\sigma_b=\epsilon_o [ (\epsilon_{rin}-1)E_{inn} - (\epsilon_{rout}-1) 
E_{outn}]
\end{equation}
where $E_{inn}$ is the normal component of the electric field
just inside the surface.  This is equivalent to:
\begin{equation}
\label{8.4}
\sigma_b=D_{inn} - D_{outn} + \epsilon_o (E_{outn}-E_{inn})
\end{equation}

The absence of free charge means that $D_{inn}=D_{outn}$.
Therefore
\begin{equation}
\label{8.5}
\sigma_b= \epsilon_o (E_{outn}-E_{inn})
\end{equation}

This provides a formula for dipole moment in terms of the electric field:
\begin{equation}
\label{8.6}
{\bf p}=\epsilon_o \int{\bf r}({\bf E}_{out}-{\bf E}_{in}) \cdot d{\bf A}
\end{equation}

The electric field at great distances is ${\bf E}_{inc}$. It
can also be noted that the electric field in and near the
object in this situation depends only on
$\epsilon_{in}/\epsilon_{out}$.
To see why, note that $\nabla \cdot {\bf D} = 0$. Therefore,
$\nabla \cdot (\epsilon({\bf r}) {\bf E} ({\bf r}))=0$.
Consequently, a given electric field still solves the
boundary value problem if both $\epsilon_{in}$ and
$\epsilon_{out}$ are multiplied by the same constant.

Because of Eq.~(\ref{8.6}), the same also holds true for the
dipole moment:
\begin{equation}
\label{8.7}
{\bf p}=f({\bf E}_{inc}, \frac{\epsilon_{in}}{\epsilon_{out}} )
\end{equation}
where $f()$ is some (non-scalar) function that also depends on
the size, shape, and orientation of the object.

Thus, if the dipole moment is theoretically calculable for the
situation of a homogeneous object sitting in a vacuum,
Eq.~(\ref{8.7}) can
be used to find the dipole moment when permittivities inside and
outside are multiplied by the same constant. The dipole moment
will be the same, in fact.

\providecommand{\href}[2]{#2}\begingroup\raggedright\endgroup
\end{document}